\documentstyle[emulateapj,graphicx]{article}
\begin{document}

\title{External shock model for the large-scale, relativistic X-ray jets from the
microquasar XTE J1550-564}

\author{X. Y. Wang,  Z. G. Dai  and T. Lu}
\affil{Department of Astronomy,
Nanjing University, Nanjing 210093, P.R.China
}

\begin{abstract}
Large-scale, decelerating, relativistic X-ray jets due to material
ejected from the black-hole candidate X-ray transient and
microquasar XTE J1550-564 has been recently discovered with {\em
Chandra} by Corbel et al. (2002). We find that the dynamical
evolution of the eastern jet at the late time is consistent with
the well-known Sedov evolutionary phase.  A trans-relativistic
external shock dynamic model by analogy with the evolution of
gamma-ray burst remnants, is shown to be able to fit the
observation data reasonably well. The inferred interstellar medium
density around the source is well below the canonical value
$n_{\rm ISM}\sim1~ {\rm cm^{-3}}$. We find that the emission from
the continuously shocked interstellar medium (forward shock
region) decays too slowly to be a viable mechanism for the eastern
X-ray jet. However, the rapidly fading X-ray emission can be
interpreted as synchrotron radiation from the non-thermal
electrons in the adiabatically expanding ejecta. These electrons
were accelerated by the reverse shock (moving back into the
ejecta) which becomes important  when the inertia of  the swept
external matter leads to an appreciable slowing down of the
original ejecta. To ensure the dominance of the emission from the
shocked ejecta over that from the forward shock region during the
period of the observations, the magnetic field and electron energy
fractions in the forward shock region must be far below
equipartition. Future continuous, follow-up multi-wavelength
observations of  new ejection events from microquasars up to the
significant deceleration phase should provide more valuable
insight into the nature of the interaction between the jets and
external medium.

\end{abstract}

\keywords{stars:individual (XTE J1550-564)--- gamma rays:
bursts---ISM: jets and outflows--- radiation mechanisms:
non-thermal}

\section{Introduction}

Black-hole X-ray binaries with relativistic jets resemble, on a
much smaller scale, many of the phenomena seen in quasars and are
therefore called microquasars (Mirabel \& Radriguez 1999).
Apparently superluminal radio jets are observed from at least the
two best known microquasars GRS 1915+105 ( Mirabel \& Rodriguez
1994; Fender et al. 1999) and GRO J1655-40 (Tingay et al. 1995;
Hijellming \& Rupen 1995), with the actual jet velocities greater
than 0.9c. Their motions are consistent with being purely
ballistic, up to a projected separation of 0.08pc from the compact
object on a maximum time scale of 4 months after the ejection for
GRS 1915+105. With the help of {\em Chandra} Observatory, Corbel
et al. (2002) recently discovered large-scale (with projected
separation more than a half pc from the compact object),
relativistically moving and decelerating radio and X-ray emitting
jets from the microquasar XTE J1550-564. It is the first time that
an X-ray jet proper-motion measurements spanning several years are
obtained for accretion-powered Galactic sources. Hence, XTE
J1550-564 provides a good opportunity to study the dynamical
evolution of relativistic jets.

In this paper, we propose that the dynamical evolution and
radiation of the large-scale X-ray jets from XTE J1550-564 can be
understood as the interaction between the jet and the surrounding
interstellar medium (ISM), quite similar to the external shock
model for afterglows of gamma-ray bursts (GRBs) (Rees \&
M\'{e}sz\'{a}ros 1992; M\'{e}sz\'{a}ros \& Rees 1997). The main
differences lie in the initial Lorentz factor and the kinetic
energy of the jets. The microquasar jets are inferred to be at
least mildly relativistic with initial Lorentz factor
$\Gamma_0\sim 2-5$ (Mirabel \& Radriguez 1999){\footnote{Recently,
Fender (2003) argued that the two-sided jet proper motions
observed from microquasars can only allow placing lower limits on
the initial Lorentz factors. }}, in contrast to the
ultra-relativistic jets ($\Gamma_0>100$) in GRBs. When the
relativistic ejecta from the microquasar is significantly
decelerated by the ISM,
 a relativistic {\em forward} shock expands into the ISM and a {\em reverse}
 shock moves into and heats the original ejecta. The shocked ambient and
 ejecta materials are in pressure balance and separated by a contact discontinuity.
The forward shock continuously heats fresh ISM  and accelerates
electrons, while the reverse shock operates only once and after
that the shocked gas in the ejecta expands and cools
adiabatically. As more and more ISM matter are swept-up, the
ejecta and the shocked ISM
--- we shall call them 'jet'--- should be decelerated more and more
and finally transit to the non-relativistic motion phase. We find
that this dynamic model can fit well the observed proper-motion
evolution of the large scale eastern  X-ray jet from XTE
J1550-564.

In both shocks,   the kinetic energy are converted to the internal
energy. In the standard theory of GRB afterglows, it is assumed
that shocked electrons and the magnetic field acquire   constant
fractions of the total shock energy, denoted by $\epsilon_e$ and
$\epsilon_B$ for electrons and magnetic field respectively. The
inferred values from a few afterglows are in the range:
$\epsilon_e\simeq 0.1-0.6$ and $\epsilon_B\simeq10^{-6}-0.1$ ({
e.g. Granot et al. 1999; Wijers \& Galama 1999; Panaitescu \&
Kumar 2002}). If we assume that the same processes also occur in
the large-scale, decelerating jet we consider here, we are able to
know the emission from the forward shock and reverse shock
regions. We find that forward shock emission decays too slowly to
be consistent with the X-ray emission from the eastern jet of XTE
J1550-564, but the emission from the post-shocked adiabatically
expanding ejecta can give a reasonable fit.

First, we give a brief review of the observations of the large
scale jets from XTE J1550-564 in section 2. We present the dynamic
model  fit in section 3 and interpret the radiation in section 4.
Finally, we give conclusions and discussions.

\section{Observations of the large-scale jets from XTE J1550-564}
The X-ray transient XTE J1550-564 was discovered by the All-Sky
Monitor on aborad the {\it Rossi X-ray Timing Explorer} on 7
September 1998 (Smith 1998). Optical observations of the source in
quiescence showed that the mass of the compact object is near
$10M_\odot$, indicating that the compact object is a black hole
and revealed that the binary companion to be low-mass star
(Hannikainen et al. 2001). The distance ($d$) to the source is
constrained to be in the range 2.8-7.6 kpc with a favored value of
5.3 kpc (Orosz et al. 2002). Soon after the discovery of the
source, an extremely strong X-ray flare was observed on 20
September 1998 (Sobczak et al. 2000; Homan et al. 2001), and radio
jets with apparent superluminal velocities (the initial proper
motion was greater than $57 {\rm ~mas~ day^{-1}}$) was observed
beginning 24 September 1998 (Hannikainen et al. 2001). During the
2002 X-ray outburst, radio observations were made with the {\it
Australia Telescope Compact Array (ATCA)}. The detection of the
large-scale radio jet $\sim 22 ~{\rm arc sec}$ to the west of XTE
J1550-564 led to a re-analysis of the archival {\em Chandra} data
and discovery of an X-ray jet to the east of XTE J1550-564 (Corbel
et al. 2002; Tomsick et al. 2003). It is thought that both jets
are connected with the 20 September 1998 ejection event, based on
the detection of superluminal jets following an extremely large
X-ray flare and the absence of any other X-ray flare of similar
magnitude in continual X-ray monitoring from 1996 to 2002.

According to the archival {\em Chandra} data, the field of view of
XTE J1550-564 was imaged by {\em Chandra} on 9 June, 21 August,
and 11 September 2000. These observations present for the first
time the proper-motion measurement for an X-ray jet from
microquasars, which show direct evidence for gradual deceleration.
Radio observations with ATCA show a decaying and moving radio
source consistent with the position of the eastern X-ray jet.
Remarkably, the overall radio flux detected on 1 June 2002 for the
eastern jet is consistent with an extrapolation of the X-ray
spectrum with a single  power law of spectral index of
$\alpha\simeq -0.6$ ($F_\nu\propto \nu^{\alpha}$) (Tomsick et al.
2003), which provides evidence for a synchrotron X-ray emission
mechanism{\footnote{ The broadband spectral energy distribution of
the western jet around 11 March 2002, which is consistent with a
single  power law of spectral index of $-0.660\pm0.005$,
strengthened the synchrotron radiation origin of the X-ray
emission (Corbel et al. 2002) .}}. The field of view of XTE
J1550-564  was further observed by {\em Chandra} on 11 March 2002.
We summarize the observations for both the eastern and western
jets in Table 1.

For the western jet, Kaaret et al. (2003) also examined the
archival {\em Chandra} data on XTE J1550-564 at June, August and
September 2000, but found no evidence for X-ray emission in any of
the archival observations with an upper bound on the absorbed flux
of $1.1\times 10^{-14}{\rm erg cm^{-2} s^{-1}}$. The facts that no
X-ray emission detected from the western jet during 2000 and that
the eastern source apparently moves faster than the western one
are consistent with the interpretation in which the eastern jet is
the approaching one and the western jet is the receding one. The
brightening of the western jet at the late time  argues against
symmetric jet propagation and may reflect the non-uniformity in
the ISM.

\section {The dynamic model}
As we only know the proper motion measurements and the light curve
of the eastern jet, we focus on modelling the dynamical evolution
and radiation of this jet.

We envision a beamed outflow with kinetic energy $E_0$ and Lorentz
factor $\Gamma_0$ ejected from the microquasar XTE J1550-564
expands conically with a half opening  angle $\theta_j$ into the
ambient medium with a constant number density $n$. The interaction
between the relativistic ejecta and the surrounding medium is
analogous to  GRB external shock, but with quite different $E_0$
and $\Gamma_0$. The supersonic motion of the ejecta should drive a
blast wave propagating into the ISM. We shall assume that the
radiation loss of the shock wave is  a negligible fraction of the
total energy, so the dynamic is adiabatic throughout. From the
view of the energy conservation, the dynamic equation can be
simplified as (see also Huang, Dai \& Lu 1999)
\begin{equation}
(\Gamma-1)M_0 c^2 + \sigma (\Gamma_{\rm sh}^2-1)m_{\rm sw}
c^2=E_0,
\end{equation}
where $\Gamma$ and $\Gamma_{\rm sh}$ are the Lorentz factors of
the jet material and the shock front respectively, $m_{\rm
sw}=(4/3)\pi R^3 m_p n (\theta_j^2/4)$ is the mass of the swept-up
ISM (where $R$ is the shock radius, $m_p$ is the mass of the
proton) and $M_0$ is the mass of the original ejecta. The first
term on the left of the equation is the kinetic energy of the
ejecta and the second term  is the internal energy of the shock.
For ultra-relativistic shocks $\sigma=6/17$, while $\sigma=0.73$
for non-relativistic shocks (Blandford \& Mckee 1976)
{\footnote{Strictly speaking, these values are corresponding to
point explosions with shock structure described by the similarity
solution. }}. For simplicity, we approximate this term as
$0.7(\Gamma^2-1)m_{\rm sw}c^2$ in the following calculation (Note
that at observation time the jet has transited to the
sub-relativistic motion phase and that a slight difference of
$\sigma$ may reflect it in the value of $E_0$, which is a free
parameter here).

The kinematic equation of the approaching jet is
\begin{equation}
\frac{dR}{dt}=\frac{\beta(\Gamma)c}{1-\beta(\Gamma){\rm
cos}\theta},
\end{equation}
where $v=\beta c$  is the bulk velocity of the jet with
$\beta(\Gamma)=(1-\Gamma^{-2})^{1/2}$, $t$ is the observer time
and $\theta$ is the jet inclination angle to the line of sight.

Given the initial condition at $t=0$,  which is chosen to be
$R_0=10^7 {\rm cm}$ and $\Gamma_0=3$, the two equations can be
solved and we can get the relation between the proper motion $\mu$
($\mu=R{\rm sin}\theta/d=R{\rm sin}\theta/5.3{\rm kpc}$) and time
$t$.

We find that the following combination of the parameters fits the
observed proper-motion data reasonably well: $E_0=3.6\times10^{44}
{\rm erg}$, $n=1.5\times10^{-4}{\rm cm^{-3}}$,
$\theta_j=1.5^\circ$ and
$\theta=50^\circ$.{\footnote{$\theta=50^\circ$ and $\Gamma_0=3$
are consistent with the initial proper motion $>57~{\rm
mas~day^{-1}}$ for $d=5.3{\rm kpc}$. }} The model fit is plotted
in Figure 1 as the solid line. The late time behavior approaches
the well-known Sedov solution $R\propto t^{2/5}$ (see the dashed
line in Fig. 1). This is expected since the second term on the
left of Eq.(1) becomes dominant at the late time, so $\beta^2
R^3={\rm constant}$. In addition, $\frac{dR}{dt}\simeq \beta c$
when $\beta {\rm cos}\theta\ll1$. The earlier phase can be
regarded as the transition regime from the mildly relativistic
motion to the non-relativistic motion.  By comparison, we also
plot the $R\propto t^{0.5}$ and $R\propto t^{0.3}$ cases in Figure
2, which show clear deviations from the data.

The inferred value of the ISM density is surprisingly low. Such
low densities with $n\la 10^{-3}{\rm cm^{-3}}$ have been inferred
around another two microquasars GRS 1915+105 and GRO J1655-40 by
Heinz (2002) from the fact that jets move with constant velocities
up to distance $\ga0.04$ pc. As argued by Heinz (2002), this
implies that either the sources are located in regions occupied by
the hot ISM phase or  previous activities of the jets have created
evacuated bubbles around the sources.

\section{The radiation model}
\subsection {The forward shock emission}
First, we try to fit the X-ray emission of the eastern jet using
the forward shock model, the mechanism believed to be responsible
for GRB afterglows. In the standard picture of GRBs, an afterglow
is generally believed to be produced by the synchrotron radiation
or inverse Compton emission of the shock-accelerated electrons in
an ultra-relativistic shock wave expanding into the ambient
medium. As more and more ambient matter is swept up, the shock
gradually decelerates while the emission from such a shock fades
down. The microquasar jet is similar to the GRB remnant at the
time that its Lorentz factor has decreased to $\Gamma\sim 1-3$,
usually months to years after the burst. So, we expect that
similar emission processes should also occur in the case of the
microquasar decelerating jet.

If the distribution of the shock-accelerated electrons takes a
power-law form with the number density given by
$n(\gamma_e)d\gamma_e=K \gamma_e^{-p}d\gamma_e$ for
$\gamma_m<\gamma_e<\gamma_M$, the volume emissivity at the
frequency $\nu'$ in the comoving frame of the shocked gas is
(Rybicki \& Lightman 1979)
\begin{equation}
j_{\nu'}=\frac{\sqrt{3}q^3}{2 m_e c^2}\left(\frac{4\pi m_e c
\nu'}{3q}\right)^{\frac{1-p}{2}}B_\bot^{\frac{P+1}{2}}K
F_1(\nu',\nu'_m,\nu'_M),
\end{equation}
where $q$ and $m_e$ are respectively the charge and mass of the
electron, $B_\bot$ is the strength of the component of magnetic
field perpendicular to the electron velocity, $\nu'_m$ and
$\nu'_M$ are the characteristic frequencies for electrons with
$\gamma_m$ and $\gamma_M$ respectively, and
\begin{equation}
F_1(\nu',\nu'_m,\nu'_M)=\int^{\nu'/\nu'_m}_{\nu'/\nu'_M}F(x)x^{(p-3)/2}dx
\end{equation}
with $F(x)=x\int^{+\infty}_x K_{5/3}(t)dt$ ($K_{5/3}(t)$ is the
Bessel function).

The physical quantities in the pre-shock  and post-shock ISM are
connected by the jump conditions:
\begin{equation}
n'=\frac{\hat{\gamma}\Gamma+1}{\hat\gamma-1}n,~~~e'=\frac{\hat{\gamma}\Gamma+1}
{\hat{\gamma}-1}(\Gamma-1)nm_p{c^2},
\end{equation}
\begin{equation}
\Gamma_{\rm sh}
^2=\frac{(\Gamma+1)[\hat{\gamma}(\Gamma-1)+1]^2}{\hat{\gamma}(2-
\hat{\gamma})(\Gamma-1)+2},
\end{equation}
where $e'$ and $n'$ are the energy and the number densities of the
shocked gas in its comoving frame and $\hat{\gamma}$ is the
adiabatic index, which equals $4/3$ for ultra-relativistic shocks
and $5/3$ for sub-relativistic shocks. A simple interpolation
between these two limits $\hat{\gamma}={(4\Gamma+1)}/{3\Gamma}$
gives a valid approximation for trans-relativistic shocks (Dai,
Huang \& Lu 1999).

Assuming that shocked electrons and the magnetic field acquire
constant fractions ($\epsilon_e$ and $\epsilon_B$) of the total
shock energy, we get
\begin{equation}
\gamma_m=\epsilon_e\frac{p-2}{p-1}\frac{m_p}{m_e}(\Gamma-1),
B_\bot=\sqrt{8\pi \epsilon_B e'}
\end{equation}
and
\begin{equation}
K=(p-1)n'\gamma_m^{p-1}
\end{equation}
{ for $p>2$}.{\footnote{ The expressions of $\gamma_m$ and $K$ are
different if $p<2$ (see e.g. Dai \& Cheng 2001).}} It is
reasonable to believe that $\nu'_M$ is well above the X-ray band
throughout the observations, because the forward shock
continuously heats fresh ISM and accelerates electrons. The
observer frequency $\nu$ relates to the frequency $\nu'$ in the
comoving frame by $\nu=D\nu'$, where $D={1}/{\Gamma(1-\beta {\rm
cos}\theta)}$ is the Doppler factor. The observed flux density at
$\nu$ and the X-ray flux in the band 0.3-8keV are respectively
given by
\begin{equation}
F_\nu=\frac{\theta_j^2}{4}(\frac{R}{d})^2\Delta R D^3 j_{\nu'};~~~
F(0.3-8{\rm keV})=\int^{\nu_2}_{\nu_1}F_\nu d\nu,
\end{equation}
where $\Delta R$ is the width of the shock region and is assumed
to be $\Delta R=R/10$ in the calculation. The model fit to the
light curve of the eastern jet is presented in Figure 3. Clearly,
the model light curve decays too slowly to fit the observed data.
This is expected from the following analytic study. For
$\nu'_m\ll\nu'\ll\nu'_M$, $F_{\nu}\propto D^3
B_\bot^{(p+1)/2}KR^3\propto D^3
B_\bot^{(p+1)/2}n'\gamma_m^{p-1}R^3$. When $\beta\ll1$,
$\beta\propto t^{-3/5}$, $R\propto t^{2/5}$, $B\propto \beta$,
$\gamma_m\propto\beta^2$ and $n'=4n={\rm constant}$. So, we get
$F_\nu\propto t^{-(15p-21)/10}\propto t^{-1.2}$ for $p=2.2$. We
also consider the case in which the jet spreads laterally  with
the sound velocity as discussed in GRBs(Rhoads 1999), but find
that it makes little difference.

\subsection{The reverse shock emission}
Another probable emitting region for X-rays is the adiabatically
expanding  ejecta itself. A reverse shock wave that moves back
through the original ejecta becomes important when the swept-up
matter mass equals $1/\Gamma_0$ of the ejecta mass. The reverse
shock accelerates electrons in the ejecta and may amplify the
original magnetic field  to be close to equipartition. The
emission from the non-thermal ($N(\gamma_e)\propto \gamma_e^{-p}$)
relativistic electrons in the adiabatically expanding ejecta with
radius $R$ is described by the van der Laan (1966) model, where
the flux density is given by $F_\nu\propto R^{-2p}\nu^{(1-p)/2}$
in the optically-thin regime. So, when $\beta\ll1$, $F_\nu\propto
t^{-4p/5}\propto t^{-1.76}$ for $p=2.2$. This decay is still too
slow to fit the observed light curve of the X-ray jet for which
the power-law fit gives a decay index of $-3.7\pm0.7$ (Kaaret et
al. 2003). However, different from the continuous forward shock,
the reverse shock operates only once, so the electrons, including
those with the maximum energy  in the ejecta, all cool
adiabatically. So it is likely that, at some time, the
characteristic frequency of the electrons with the maximum energy
$\gamma_M m_e c^2$ may fall close to the X-ray band and the X-ray
flux would decay quite rapidly since then.

The maximum energy of the power-law distribution electrons just
after the reverse shock crosses the ejecta is determined by the
shock acceleration process, which is however not well understood.
If the electrons  in the high energy end of the power law cool
faster than the dynamical time scale,
  the real  maximum energy of the electrons in the  ejecta  is
limited by  the synchrotron cooling timescale; electrons with
energy greater than this energy would have cooled down within the
dynamical timescale. So, $E_M^0=\min({E_{M,acc},E_{M,cool}})$,
where $E_{M,acc}$ and $E_{M,cool}$ denote the maximum energies
allowed by the shock acceleration process and the cooling process
respectively, the superscript 0 in $E_M^0$ denotes the value at
 $t_0$--- the time when the reverse shock heats the ejecta.
If the synchrotron radiation dominates the cooling of the
electrons, $E_{M,cool}=6\pi m_e^2 c^3/(\sigma_T B_\bot^2 t'_0)$,
where $t'_0=t_0/D(t_0)$ is the dynamical time in the comoving
frame, $D(t_0)$ is the Doppler factor of the X-ray jet at the time
$t_0$, $\sigma_T$ is the Thomson cross section.

The physical quantities in the adiabatically expanding ejecta with
radius $R$ relate  with their initial value at $t_0$ by ( van der
Laan 1966)
\begin{equation}
\gamma_m=\gamma_m(t_0)\frac{R_0}{R},~~~~~~
\gamma_M=\gamma_M(t_0)\frac{R_0}{R};
\end{equation}
\begin{equation}
K=K(t_0)\left(\frac{R}{R_0}\right)^{-(2+p)},
B_\bot=B_\bot(t_0)\left(\frac{R}{R_0}\right)^{-2}
\end{equation}
if  the synchrotron emission cooling is negligible,  where $R_0$
is the radius of the ejecta at time $t_0$ and
$\gamma_M(t_0)=E_M^0/m_e c^2$. These relations are derived from
the assumption that total number of the electrons is conserved and
that the magnetic field is frozen to the plasma
fluid{\footnote{The same relation for the magnetic field holds if
we assume that the energy in the magnetic field constitutes a
constant fraction of the internal energy in the ejecta during the
whole adiabatically expanding phase. }}.

From the dynamic model in section 3, we get  $t_0=121$ days and
$R_0=0.71\times10^{18}{\rm cm}$ for the  eastern X-ray jet from
XTE J1550-564. Given the initial condition for $K(t_0)$,
$B_\bot(t_0)$, $\gamma_m(t_0)$ and $\gamma_M(t_0)$, we can obtain
the model light curves of the emissions from  the expanding ejecta
using Eqs.(3), (4), (9) and $R(t)$. In the calculation, we have
assumed $\gamma_M(t_0)=E_{M,cool}/m_e c^2=6\pi m_e c/(\sigma_T
B_\bot(t_0)^2 t'_0)$. We find that the following combination of
the initial values  can fit the flux data well (see Figure 4):
$B_\bot(t_0)=0.5 {\rm~ m G}$ and $K(t_0)=0.8~ {\rm cm^{-3}}$ and
$\gamma_m(t_0)=100$.{\footnote{The model result is very
insensitive to the value of $\gamma_m(t_0)$}} For these values,
$\gamma_M(t_0)=3.55\times10^8$.

 Noting that
$K=\epsilon_e(p-2)e'\gamma_m^{p-2}$ and
$B_\bot=\sqrt{8\pi\epsilon_B e'}$, the above value for $K(t_0)$
and $B_\bot(t_0)$ imply that the equipartition parameters  for
energies in electrons and magnetic field in the reverse shock are
$\epsilon_e=0.6$ and $\epsilon_B=0.44\times10^{-2}$ respectively,
if the internal energy density in  the reverse shock is equal to
that in the forward shock region at time $t_0$. Surprisingly,
these equipartition values  are  close to the typical values
inferred for GRB afterglows (Granot et al. 1999, Wijers \& Galama
1999; Wang, Dai \& Lu 2000). { Please also note that here the
emitting electrons are in slow cooling regime as
$\nu_m(t_0)\sim10^7 {\rm Hz}\ll\nu_c$, where $\nu_c$ is the
cooling frequency, and that at the X-ray band the synchrotron
emission dominates over the inverse Compton emission (Sari \& Esin
2001; Panaitescu \& Kumar 2000).}

We further calculate the model spectrum for the eastern jet on 1
June 2000 using the above parameter values and plot the fit of the
observed data in Figure 5. Clearly the model fits quite well the
energy spectrum on 1 June 2000, 621 days after the ejection of the
eastern jet from XTE J1550-564. At this time, the characteristic
frequency of the electrons with $\gamma_M$ is just near the X-ray
band, so the X-ray spectrum doesn't become steeper. The model
predicts that there should be significant steepness of X-ray
spectrum at March 2002, but the relatively few counts do not give
a reliable spectral index (Kaaret et al. 2003).

We have shown that the  emission from the shocked ejecta provides
a viable mechanism for the rapidly decaying X-ray flux from the
eastern jet of XTE J1550-564. { To guarantee the dominance of the
emission from the shocked ejecta  over that from the forward shock
region during the period of the observations, the forward shock
emission should be below the upper limit of the observation made
on June 19 2002. Fig.3 tells us that the forward shock emission
for $\epsilon_e=0.1$ and $\epsilon_B=10^{-4}$ is inferred to be
below the limit. Because $F_\nu\propto
\gamma_m^{p-1}B^{(p+1)/2}\propto\epsilon_e^{p-1}\epsilon_B^{(p+1)/4}$,
the magnetic field and electron energy fractions in the forward
shock region must satisfy (see Fig. 3) }
\begin{equation}
\left(\frac{\epsilon_e}{0.1}\right)^{p-1}\left(\frac{\epsilon_B}{10^{-4}}\right)
^{(p+1)/4}\la1.
\end{equation}

The  high value for $\epsilon_B$ in the reverse shock region
relative to that in the forward shock region can be accounted for
if the ejecta has already been magnetized before the further shock
compression{\footnote{The high value of $\epsilon_B$ inferred for
GRB afterglows can, however, be attributed to particular
environment (e.g. K\"{o}nigl \& Granot 2002) around the burst or a
particular magnetic field amplification mechanism (e.g. Thompson
\& Madau 2000).}}. This is reasonable as    the ejecta from
microquasars are suggested to originate from the inner accretion
disks (Mirabel \& Rodriguez 1999).  A pair-rich outflow from the
microquasar probably account for the comparatively large value of
$\epsilon_e$ in the reverse shock. Future observations of
large-scale jets from microquasars may provide better
understanding of the shock physics and physical condition in
relativistic jets.

\section{Conclusions and Discussions}
It is believed that relativistic jets exist in accreting systems
ranging from Galactic X-ray binaries, gamma-ray bursts and active
galactic nuclei (AGNs). Jets in Galactic X-ray binaries such as
XTEJ1550-564 evolves much more rapidly than AGN jets and therefore
offer a good opportunity to study the dynamical evolution of
relativistic jets on time scales inaccessible for AGNs. Although
afterglows in GRBs evolves also rapidly, their cosmological
distances make the direct measurements of the proper-motion
impossible and their dynamics can be studies only indirectly.

The discovery of the extended radio and X-ray emission from the
microquasar XTE J550-564 (Corber et al. 2002; Tomsick et al. 2003;
Kaaret et al. 2003) represents the first detection of large-scale
relativistic jets from a Galactic black hole candidate in both
radio and X-rays. These large-scale jets appear to arise from a
relatively brief ejection event and, therefore, offer a unique
opportunity to study the large-scale evolution of an impulsive
jet. We find that the dynamical evolution of the observed eastern
jet  from the  XTE J550-564 is consistent with the well-known
Sedov evolutionary phase, during which the energy in the jet is
conserved and $R\propto t^{2/5}$. The apparent superluminal motion
observed at the very early epoch implies that the initial motion
of the jet is at least mildly relativistic. As more and more ISM
matter is swept up, the jet decelerates and finally transits to
the non-relativistic phase. A trans-relativistic external shock
dynamical model is shown to be able to fit the observed proper
motion data reasonably well. The inferred ISM density around the
jet is $n\sim1.5\times10^{-4}{\rm cm^{-3}}$, well below the
canonical value. Such low ISM density gains support from the
inferred value $n\la 10^{-3}{\rm cm^{-3}}$ around another two
microquasars GRS 1915+105 and GRO J1655-40 by Heinz (2002), from
the fact that jets move with constant velocities (i.e. no slowing
down) up to distance $\ga0.04$ pc. As suggested by Heinz (2002),
this implies either that  the sources are located in regions
occupied by the hot ISM phase or that previous frequent activities
of the jets have created evacuated bubbles around the sources.

We first try to fit the X-ray light curve of the eastern jet with
the emission from the shocked ISM. However, it is found that this
predicts a decay too  slow to fit the observations. The model
predicts $F_\nu\propto t^{-(15p-21)/10}\sim t^{-1.2}$ during the
non-relativistic phase, while the power law  fit of the X-ray flux
data gives $F_\nu\propto t^{-3.7\pm0.7}$ (Kaaret et al. 2003). We
then turn to another likely emission region---the adiabatically
expanding ejecta heated by the reverse shock, quite similar to the
mechanism suggested to be responsible for the optical flash and
radio flare from GRB990123 (Sari \& Piran 1999). Different from
the shocked ISM, all electrons in the ejecta cool by adiabatic
expansion, so the maximum energy of the electrons in the ejecta
decreases as well{\footnote{ However, for the forward shock, there
are always new electrons being shocked. $\gamma_M$ is determined
by the balance between the acceleration timescale and the cooling
timescale: $\gamma_M\propto B^{-1/2}$ (e.g. M\'{e}sz\'{a}ros et
al. 1993) . So, for the forward shock, $\gamma_M$ even increases
with time, since the magnetic field decreases with time. }}. Once
the characteristic synchrotron radiation frequency of these
electrons falls close to the X-ray band, the X-ray flux from the
ejecta should decay quite rapidly (drops exponentially with time)
since then. Using this model, we fitted both the X-ray light curve
data and the energy spectrum on 1 June 2000 of the eastern jet
successfully (see Figures 4 and 5). { One prediction we can make
here is a flattening of the light curve  at late time, once the
forward shock emission, if still above the detection limit,
overtakes the reverse shock emission.}

The western (receding) jet from XTE J1550-564 was detected in
radio and X-rays in 2002 while archival {\em Chandra} data on this
source from June, August and September 2000 only give upper
limits. The non-detection in 2000 is consistent with the deduction
that the western source is the receding jet. Unlike the smoothly
decaying eastern jet, the western jet brightens at the late time.
The brightening might be caused by the inhomogeneities in the ISM
(Kaaret et al. 2003) or internal shocks produced by a faster jet
overtaking a slower one (Kaiser et al. 2000), and need further
careful study. The western jet moved by $0.52\pm0.13$ arc sec
between 11 March and 19 June 2002 with a mean apparent speed
significantly less than the average apparent speed from 1998 to
early 2002. Interestingly, we find that the decay of the X-ray
flux of the western jet between March and June 2002 is also
consistent with that predicted by the reverse shock emission
$F_\nu\propto t^{-4/5p}\propto t^{-1.86}$ for $p=2.32$ of the
western jet.

Besides the relativistically moving, decelerating jets from XTE
J1550-564, large-scale X-ray jets and radio lobes up to $\sim 40~
{\rm arcmin}$ size have been observed from SS433 (Brinkmann et al.
1996; Dubner et al. 1998). The radiation is suggested to come from
the termination shock which results from the interaction of the
mass outflow with the nebula W50. Recently, reheating of baryonic
material in  X-ray jets of SS433 is inferred to take place within
$\sim 10^{17}{\rm cm}$ from the core, based on the observed iron
emission lines (Migliari, Fender \& M\'{e}ndez 2002). We think
that external shock is a possible mechanism for such reheating.
 As suggested for TeV neutrino emission from microquasar
jets (through internal shocks) by Levinson \& Waxman  (2002),
external shocks of microquasar jets may also accelerate the
protons in both the shocked ISM and the shocked ejecta. So, they
are also potential sources of cosmic-rays (Heinz \& Sunyaev 2002),
high-energy neutrinos and high-energy gamma-rays.

In summary, we developed a model for the dynamical evolution and
radiation of the large-scale X-ray jets from the microquasar XTE
J1550-564  analogous to  the external shock model for GRB
afterglows.  In this model, the observed jet emission is due to
interaction between the jets and external ISM. Future continuous,
follow-up multi-wavelength observations of new  ejection events
from microqusasrs up to the significant deceleration phase should
provide more valuable insights into the nature and physical
condition (e.g. shock and particle acceleration physics) of
relativistic jets.  Owing to the proximity of the Galactic X-ray
binaries, further studies on them  also offer an exciting way for
a better understanding of relativistic jets seen elsewhere in the
Universe.

{\acknowledgments We are grateful to the anonymous referee for
his/her valuable suggestions. This work was supported by the
National Natural Science Foundation of China under grants
19973003, 19825109 and 10233010, and the National 973 project. }

\newpage
{
\begin{center}
\begin{table*}[t]
\begin{center}
\caption{ Angular separations and the absorbed X-ray flux of the
eastern and western jets }
\end{center}
\begin{tabular}{|c||c||c|c||c|c|}
\hline  &  Time after X-ray& \multicolumn{2}{|c|}{Angular
Separation (arcsec)} & \multicolumn{2}{|c|}{Flux (${\rm
10^{-14}erg cm^{-2}s^{-1}}$) }\\
\raisebox{1.5ex}[0pt]{Date}& flare (days) & eastern jet & western jet & eastern jet& western jet\\
\hline \hline
 June 9 2000 & 628 & $21.3\pm0.5$ & & $20\pm6$& $<1.1$\\
\hline Aug. 21 2000 & 700 &
$22.7\pm0.5$ &  &$6.1\pm1.3$ &$<1.1$\\
\hline
Sept. 11 2000 & 720 & $23.4\pm0.5$ &&$8.2\pm1.5$ &$<1.1$ \\
\hline Mar. 11 2002 & 1265 &
$29.0\pm0.5$ &  $\sim23$& $1.1\pm0.3$&$19\pm1.0$\\
\hline   &  &
 & $\sim23+$ &&\\
\raisebox{1.5ex}[0pt]{June 19 2002 }& \raisebox{1.5ex}[0pt]{1335}
&
 & $0.52\pm0.13$ & \raisebox{1.5ex}[0pt]{$<0.3\pm0.2$}
 &\raisebox{1.5ex}[0pt]{$16\pm1.0$}\\\hline
\end{tabular}
\vskip 0.5cm
 References.--- Corbel et al. 2002; Kaaret et al. 2003; Tomsick et
 al. 2003

\end{table*}
\end{center}
}
\begin{figure*}[t]
\plotone{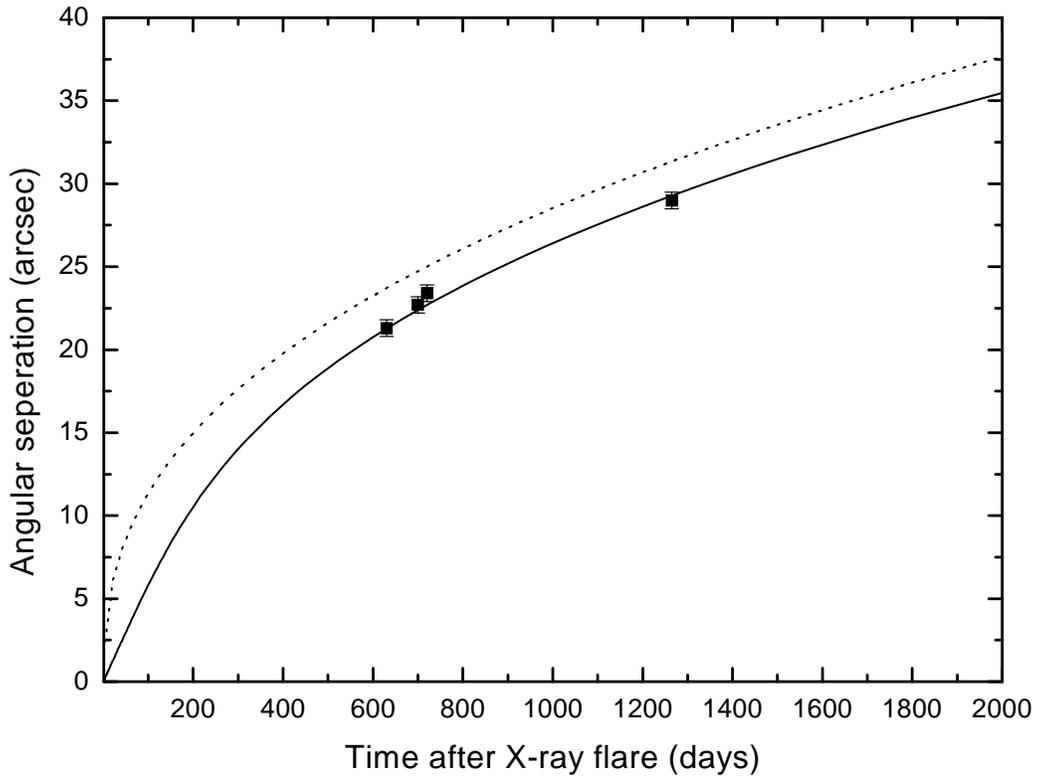} \vspace{0.3cm} \caption{Model fit to the position
of the eastern X-ray jet verse time. Observation data are taken
from  Tomsick et al. (2003) and Kaaret et al. (2003). The solid
line is the fit in terms of the trans-relativistic external shock
model. We also plot
 the function $R\propto t^{2/5}$ (dotted line) for comparison.}
\end{figure*}

\begin{figure*}[t]
\plotone{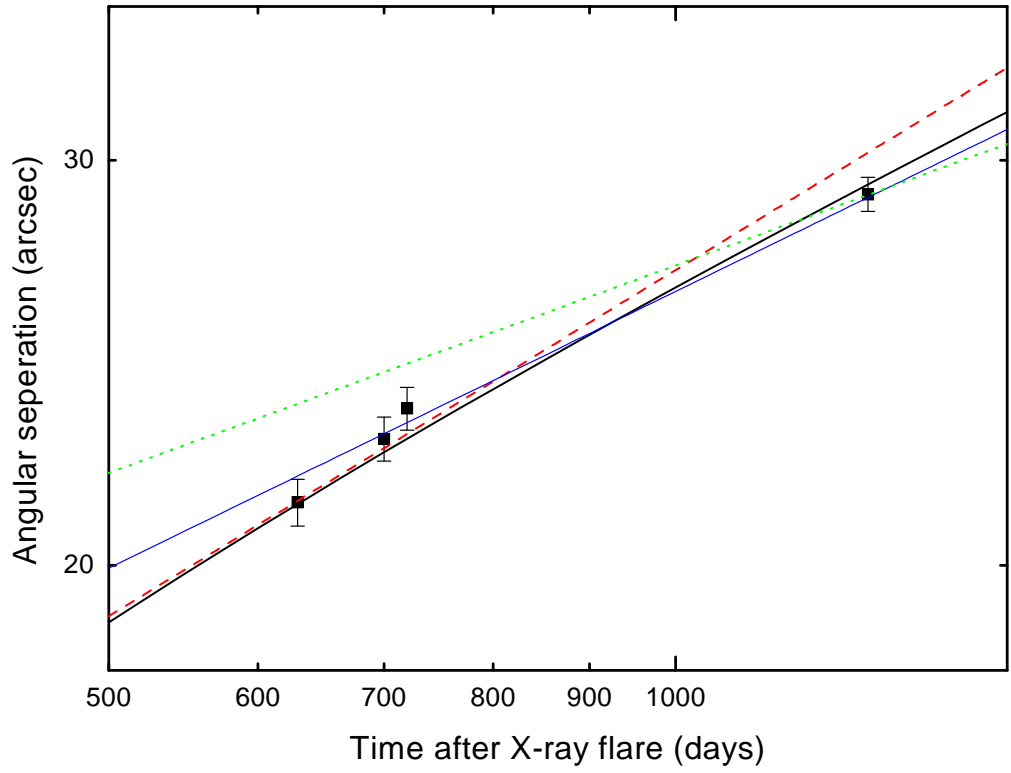} \vspace{0.3cm} \caption{Comparison of the model
fits for the  position of the eastern X-ray jet verse time. The
thick solid line is the tran-relativistic external shock model
developed in this paper. The (blue) thin  solid line, (green)
dotted line and (red) dashed line represent $R\propto t^{0.4}$,
$R\propto t^{0.3}$, $R\propto t^{0.5}$ respectively.  It shows
that the dynamical evolution of the eastern jet is consistent with
the Sedov evolutionary phase  $R\propto t^{0.4}$ at the late
time.}
\end{figure*}

\begin{figure*}[t]
\plotone{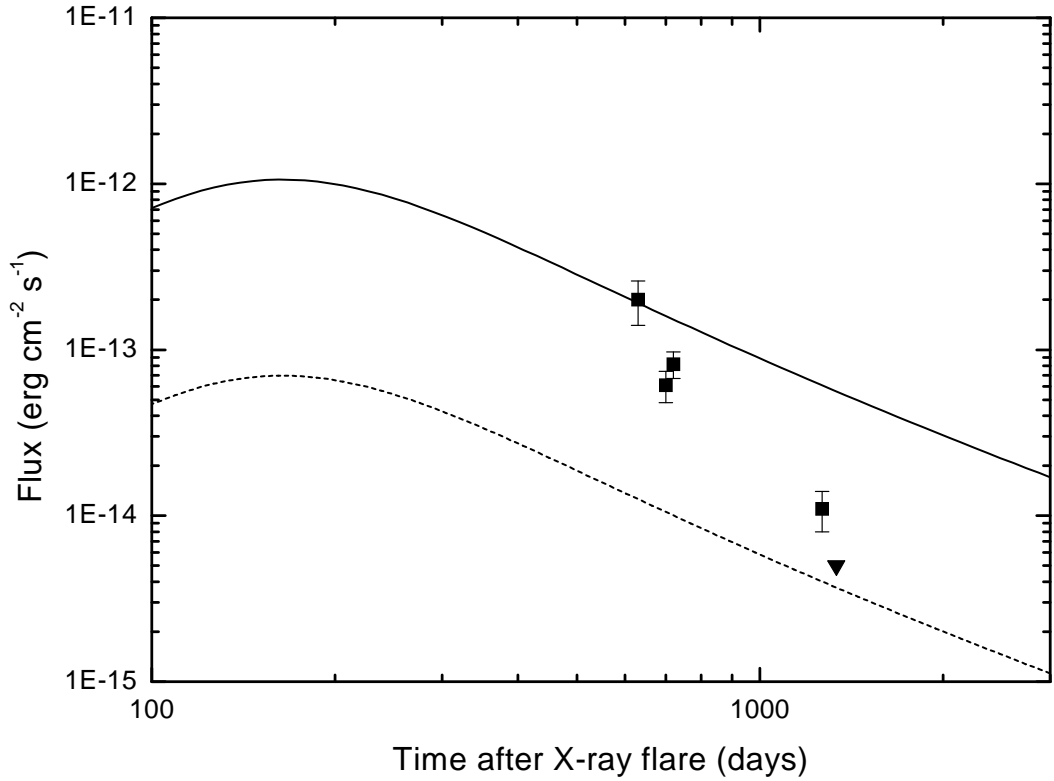} \vspace{0.3cm} \caption{Model and observed X-ray
light curves of the eastern jet. Detections and upper limits for
the non-detections, taken from Tomsick et al. (2003) and Kaaret et
al. (2003),  are indicated by the filled squares and arrows
respectively. The solid line and dotted line represent X-ray
emission from the shocked ISM (forward shock ) with different
electron ($\epsilon_e$) and magnetic field ($\epsilon_B$)
equipartition factors used. The solid and dotted lines are
corresponding to $\epsilon_e=0.1$, $\epsilon_B=0.004$ and
$\epsilon_e=0.1$, $\epsilon_B=10^{-4}$, respectively. This figure
shows that the forward shock emission decays too slowly to fit the
observations. }
\end{figure*}

\begin{figure*}[t]
\plotone{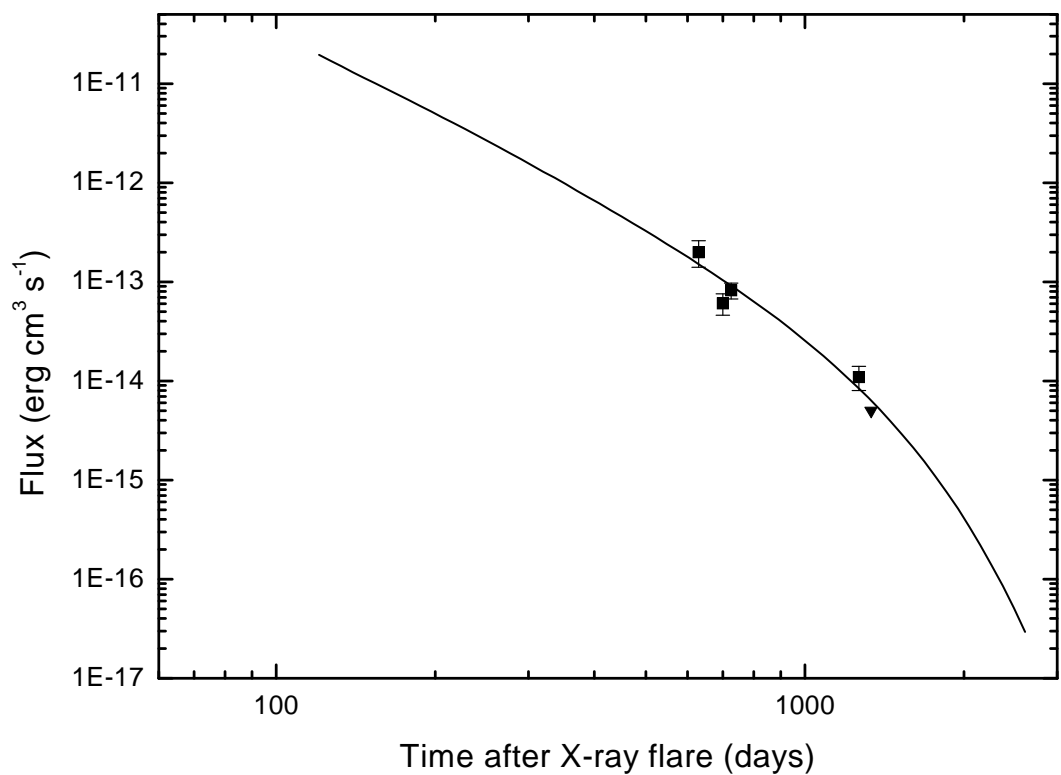} \vspace{0.3cm} \caption{Model fit to the X-ray
light curve using the synchrotron radiation from the adiabatically
expanding ejecta heated by the reverse shock.  }
\end{figure*}
\begin{figure*}[t]
\plotone{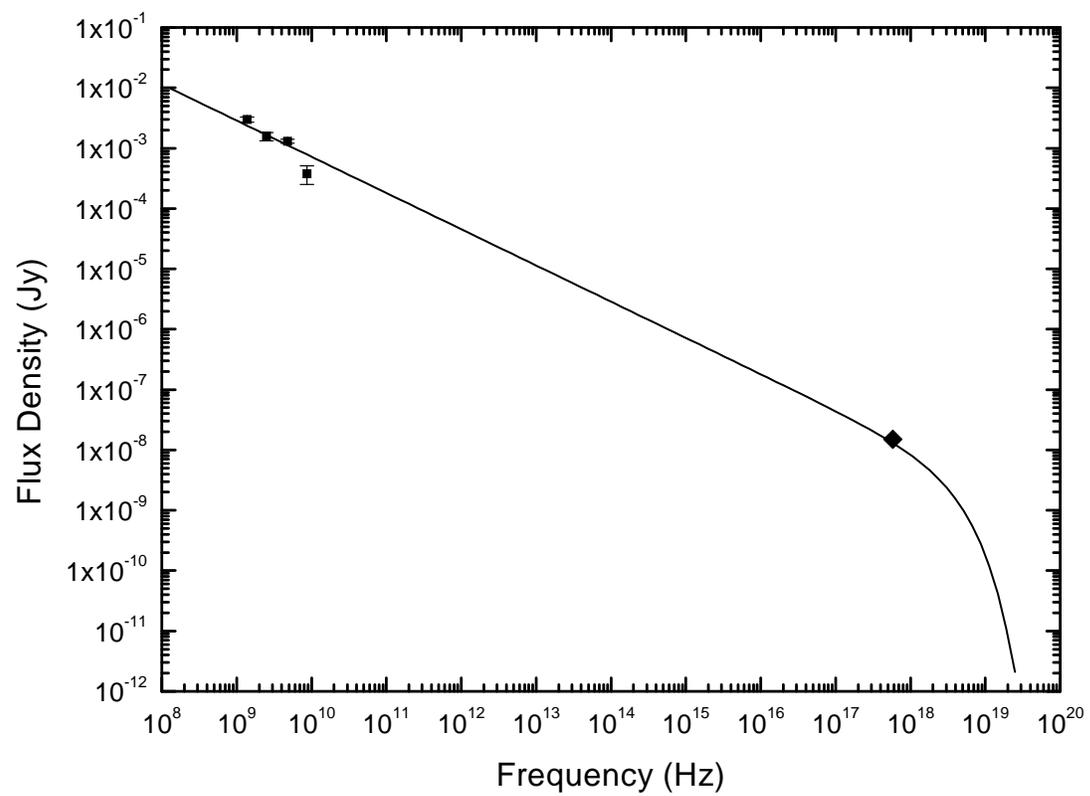} \vspace{0.3cm} \caption{Model fit to the
broadband spectrum of the eastern jet on 1 June 2000 using the
synchrotron radiation from the adiabatically expanding ejecta
heated by the reverse shock.}
\end{figure*}

\end{document}